# Forensic Taxonomy of Popular Android mHealth Apps

*Full paper*

**Abdullah Azfar**
University of South Australia
abdullah.azfar@mymail.unisa.edu.au

**Kim-Kwang Raymond Choo**
University of South Australia
raymond.choo@unisa.edu.au

**Lin Liu**
University of South Australia
lin.liu@unisa.edu.au

## Abstract

Mobile health applications (or mHealth apps, as they are commonly known) are increasingly popular with both individual end users and user groups such as physicians. Due to their ability to access, store and transmit personally identifiable and sensitive information (e.g. geolocation information and personal details), they are potentially an important source of evidentiary materials in digital investigations. In this paper, we examine 40 popular Android mHealth apps. Based on our findings, we propose a taxonomy incorporating artefacts of forensic interest to facilitate the timely collection and analysis of evidentiary materials from mobile devices involving the use of such apps. Artefacts of forensic interest recovered include user details and email addresses, chronology of user locations and food habits. We are also able to recover user credentials (e.g. user password and four-digit app login PIN number), locate user profile pictures and identify timestamp associated with the location of a user.

## Introduction

Mobile health applications (mHealth apps) such as those containing Electronic Medical Records (EMR) are gaining popularity. These apps can support health care services such as mobile telemedicine, patient monitoring, emergency response and management, and personalised monitoring. According to a recent Google report, mHealth apps are the fastest growing app category in 2014 (Research2Guidance 2014). At the time of this research, there are over 100,000 mHealth iOS and Android apps. It has also been estimated that 500 million smartphone users worldwide will use an mHealth app by 2015, and by 2018, half of the 3.4 billion smartphone and tablet users will have downloaded one or more mHealth apps (Jahns 2010; Taylor 2014; Yang and Silverman 2014). The global health and fitness mobile app market is reportedly worth about USD 4 billion at the moment, but this could increase to USD 26 billion by 2017 (Research2Guidance 2014).

The mHealth apps available on Apple app and Google play stores can be broadly categorised into patient care and monitoring; health apps for the layperson; communication, education, and research apps; and physician or student reference apps (Ozdalga et al. 2012). Most mHealth apps are designed for health and wellness management, such as cardio fitness, diet, medication adherence, women's health, strength training, stress, smoking cessation, mental health, parental and infant care, and chronic disease management (Apple 2015; Google 2015). These apps make use of the in-built features and capabilities of mobile devices to monitor user's physiological and health conditions, e.g. heart fitness apps using smartphones' cameras and flash lights to process reflections from users' fingertips to detect heart beats.

During the investigations of crimes or incidents involving mobile devices and apps, there is usually some accumulation or retention of data on the mobile device that need to be identified, preserved, analysed and presented in a court of law – a process known as mobile forensics (Barmpatsalou et al. 2013; Martini et al. 2015a; Martini et al. 2015b). Recent reviews of mobile forensics literature suggest that mobile forensics is an emerging area, but is relatively less studied than mobile security or (the "traditional") digital forensics (Barmpatsalou et al. 2013; Keith et al. 2014; La Polla et al. 2013). Despite the potential of mHealth apps in our daily life and some might refer to this as the "mobile health revolution" (Eng and Lee 2013), there has been limited advances on the forensic issues and taxonomy of mHealth apps. The importance of identifying data remnants in a forensic investigation is aptly summarised by Quick et al. (2013, p8) who noted that '[b]y determining the data remnants on client devices, we contribute to a better understanding of the types of terrestrial artifacts that are likely to remain for digital forensics practitioners and investigators'. The inability to identify relevant data in a timely fashion can adversely affect an investigation, and in the worse-case scenario, resulting in a miscarriage of justice.

A taxonomy provides an informative categorisation of data remnants in the investigation of mHealth apps. In a recent work, Plachkinova et al. (2015) proposed a security and privacy taxonomy for mHealth apps; however, they did not include forensic artefacts in their taxonomy. To the best of our knowledge, so far there has been no published forensic taxonomy for mHealth apps. This is, thus, the contribution of this paper. Similar to the approach of Plachkinova et al. (2015), we classified mHealth apps using the categorisation of Ozdalga et al. (2012). We then classified the forensic artefacts based on our study of 40 mHealth apps.

## Related Work

The security and privacy taxonomy of Plachkinova et al. (2015) for mHealth apps was tested with 38 top rated Android and iOS mHealth apps. In the taxonomy, the authors considered three dimensions, namely: the mHealth app categories (based on classification proposed by Ozdalga et al. (2012)) in one dimension, mHealth security issues in another, and mHealth privacy issues in the last dimension.

Kharrazi et al. (2012) examined 19 mHealth apps for iOS (eight apps), BlackBerry (five apps), and Android (six apps) devices, and found that seven of these apps lacked basic security measures such as password protection in user authentication. Other privacy and security challenges associated with the use of mHealth apps were also identified by He et al. (2014) and Williams (2010).

In our recent survey of 117 academic publications on mobile app security, privacy and forensics between January 2009 and August 2014, we only located 11 publications on mobile app forensics; of which four related to communication apps (Chu et al. 2013; Chu et al. 2012; Gao and Zhang 2013; Husain and Sridhar 2010); two related to social networking apps (Al Mutawa et al. 2012; Levinson et al. 2011); and five related to cloud apps (Martini and Choo 2013; Martini and Choo 2014; Quick and Choo 2013a; Quick and Choo 2013b; Quick and Choo 2014). To the best of our knowledge, there has been no research on mHealth app forensics published between January 2009 and August 2014, and at the time of this research.

## Case Study: 40 mHealth Apps

In our case study, we examined 40 popular free Android mHealth apps available on Google Play store (Google 2015). We installed and registered the apps on a Google Nexus 4 phone (Android version 5.0.1). A popular commercial forensic tool, MicroSystemation XRY (version 6.10.1), was used to extract a logical forensic image. We used a Windows 7 desktop machine to analyse the artefacts. There were individual set of experiments, one for each app. After one set of experiments was concluded, the phone was wiped prior to installing the next app.

Due to page limitation, we only discuss the findings of three apps in this section. A taxonomy proposed based on the findings of the 40 apps is shown in Figure 10 (Section 'Proposed Forensic Taxonomy for mHealth Apps') and the application of the taxonomy is provided in Appendix A.

## *Findings: Calorie Counter - MyFitnessPal*

Calorie Counter - MyFitnessPal (MyFitnessPal 2015), one of the most downloaded mHealth app (Boxall 2014), is available on iOS, Android, Blackberry and Windows devices. App users are able to access a food database containing more than four million recipes. The app has also added cardio and strength training tracker loaded with 350 exercises. Users can store their own recipes and calculate their nutritional contents, allowing them to keep track on their nutrition information such as calories, fat, protein, carbs, sugar, fiber, and cholesterol.

**Databases**

A logical extract of the MyFitnessPal app (Version 3.6.1) was conducted using XRY to obtain a logical image (i.e. forensic copy). In the context of this research, files containing information required to conduct the forensic analysis were identified - the XRY extract files and outputs such as PDF reports.

Forensic analysis of the forensic copy indicated that the files and databases generated by MyFitnessPal are stored on the internal device memory which is normally inaccessible by users. For example, the databases of MyFitnessPal are stored in `/data/data/com.myfitnesspal.android/databases` location. The recovered artefacts from the app are listed in Table 1.

| | Content | Directory | Files |
|---|---|---|---|
| 1 | Main Database | `/data/data/com.myfitnes spal.android/databases` | Myfitnesspal.db (SQLite3, 34 tables) |
| 2 | User profile pictures | `/data/data/com.myfitnes spal.android/cache/Pica sso-cache` and `/sdcard/` | stored as files and tmp_avatar_mfp~x.jpg/png in SD card |

**Table 1. Artefacts of MyFitnessPal**

The `Myfitnesspal.db` database is the app's main database comprising 34 tables. Table 2 lists the 12 tables containing artefacts of forensic interest.

| | Table name | Content |
|---|---|---|
| 1 | `user_properties` | User details including time zone, gender, date of birth and email |
| 2 | `users` | |
| 3 | `images` | User profile pictures |
| 4 | `diary_notes` | User personal notes |
| 5 - 10 | `exercise_entries; exercises; food_entries; foods; measurement_types; measurements` | User records of exercises, food habits and personal measurements |
| 11 | `last_sync_pointers` | User last synched items with the server |
| 12 | `search_history` | User food search history |

**Table 2. Tables in the `myfitnesspal.db` database of MyFitnessPal**

**User_properties Table**

The `user_properties` table in the `myfitnesspal.db` database stores user records outlined in Table 3.

| Field name | Meaning |
|------------|---------|
| `id` | A sequential value or serial number |
| `user_id` | An integer identifying a user |
| `property_name` | List of user attributes |
| `property_value` | Value for different user attributes |
| `updated_at` | Time and date when the attribute was updated |
| `last_sync_at` | Time and date when the attribute was synched with the server |

**Table 3. Structure of `user_properties` table of MyFitnessPal**

The app can be used by multiple users on a single device, and each user has a unique `user_id`, an integer value which increments by one for each user (i.e. the `user_id` for the first user is 1, 2 for the second user, ..., n for the nth user ) – see Figure 1.

The `property_name` and `property_value` fields are, probably, the most important fields for forensic investigators. The `property_name` field contains 80 attributes for each user, such as timezone identifier, gender, date of birth, country, postcode, weight, height, diary password and email address. The corresponding values for these attributes are stored in the `property_value` field. Both user attributes and their corresponding values are in plaintext, and the user information was found in the table even after the user had logged out and another user had signed in on the same device.

| id | user_id | property_name | property_value | updated_at | last_sync_at |
|-----|---------|---------------------|-------------------|---------------------|---------------------|
| 1 | 1 | timezone_identifier | A⬛⬛⬛⬛a | 2015-01-20 12:55:44 | 2015-01-20 12:56:17 |
| 6 | 1 | gender | Female | 2015-01-20 12:55:44 | 2015-01-20 12:56:17 |
| 7 | 1 | date_of_birth | 1⬛⬛⬛19 | 2015-01-20 12:55:44 | 2015-01-20 12:56:17 |
| 8 | 1 | country_name | B⬛⬛⬛l | 2015-01-20 12:55:44 | 2015-01-20 12:56:17 |
| 9 | 1 | postal_code | 123⬛ | 2015-01-20 12:55:44 | 2015-01-20 12:56:17 |
| 15 | 1 | diary_privacy_setting | password | 2015-01-20 12:55:44 | 2015-01-20 12:56:17 |
| 16 | 1 | diary_password | pa⬛⬛5 | 2015-01-20 12:55:44 | 2015-01-20 12:56:17 |
| 62 | 1 | email | ia⬛⬛⬛@gmail.com | 2015-01-20 12:55:44 | 2015-01-20 12:56:17 |
| 81 | 2 | timezone_identifier | A⬛⬛⬛e | 2015-01-18 12:10:30 | 2015-01-18 12:10:30 |
| 86 | 2 | gender | Male | 2015-01-18 12:10:30 | 2015-01-18 12:10:30 |
| 87 | 2 | date_of_birth | 1⬛⬛⬛9 | 2015-01-18 12:10:30 | 2015-01-18 12:10:30 |
| 88 | 2 | country_name | A⬛⬛⬛ | 2015-01-18 12:10:30 | 2015-01-18 12:10:30 |
| 89 | 2 | postal_code | 111⬛ | 2015-01-18 12:10:30 | 2015-01-18 12:10:30 |
| 95 | 2 | diary_privacy_setting | private | 2015-01-18 12:10:30 | 2015-01-18 12:10:30 |
| 96 | 2 | diary_password | | 2015-01-18 12:10:30 | 2015-01-18 12:10:30 |
| 142 | 2 | email | ab⬛⬛⬛r@gmail.com | 2015-01-18 12:10:30 | 2015-01-18 12:10:30 |

**Figure 1. A snapshot of the contents of `user_properties` table in MyFitnessPal**

An important feature in MyFitnessPal is the access to user diary by their friends. A user can control diary access with a password, so that only a person who is the user's friend and knows the password can access the diary. We recovered the password in plaintext in the `user_properties` table (Figure 1). An empty value for the `diary_password` field for user 2 indicates that the diary was not password-protected.

## Users table

The `users` table stores user IDs (sequential numbers), user name, encrypted passwords, and the plaintext pin codes. The pin code is a 4 digit numerical value defined by the users for a quick access to their accounts. When a user logs into his/her account with the password, the user can create this pin code. But the `users` table stores the pin codes of all the users who have logged in using the device even after logging out from the device (Figure 2).

| id | master_id | username | password | last_sync_at | pincode |
|----|-----------|----------|----------|--------------|---------|
| Filter | Filter | Filter | Filter | Filter | Filter |
| 1 | 61252726 | i⬛⬛⬛⬛1 | 94d4d38a8380ffff23cbd4c0a69df5fd145bd9e6 | 2015-01-20 12:56:17 | 2015 |
| 2 | 81074846 | ab⬛⬛⬛⬛ar | 71bc5027b0387562e0c90dbc3d30301b0ee109af | 2015-01-18 12:10:30 | 2525 |

**Figure 2. Plaintext pin code in `users` table**

## User profile pictures

The `images` table stores the URL of the profile pictures for each user who had logged into MyFitnessPal on the device. The images are associated with the `user_id` field, and the image's URL is stored even after the user has logged out. The images are stored in the cloud server and publicly accessible. A forensic investigator can obtain the URL of user profile images from the `full_image_url` field of the `images` table. The timestamps of creating and updating the profile picture are also shown in the table (Figure 3).

| user_id | image_type | filename | mbnail_image | full_image_url | created_at | updated_at |
|---------|-----------|----------|--------------|----------------|------------|------------|
| Filter | Filter | Filter | Filter | Filter | Filter | Filter |
| 2 | 1 | user/9f83/8cd3/... | http://dakd... | http://dakd0cjsv8wfa.cloudfront.net/images/photos/user/9f83/... | 2015-01-18 01:4... | 2015-01-18 01:4... |
| 1 | 1 | user/a7ab/f661... | http://dakd... | http://dakd0cjsv8wfa.cloudfront.net/images/photos/user/a7ab... | 2015-01-21 04:3... | 2015-01-21 04:3... |

**Figure 3. Profile picture URL and timestamps in the `images` table**

The user profile pictures are also stored in the SD card of the device in the location, `/sdcard/`, which is accessible by individuals with physical access to the SD card. The files are named as tmp_avatar_mfp~x.jpg or tmp_avatar_mfp~x.png, where x is a numeric value starting from 1.

## Other tables

The remaining tables containing artefacts of forensic interest are: `diary_notes`, `exercise_entries`, `exercises`, `food_entries`, `foods`, `last_sync_pointers`, `measurements_types`, `measurements` and `search_history`.

The `diary_notes` table stores the notes made by users (Figure 4). The `user_id` field identifies a user and the `note_type` field indicates whether the note is related to food (value 0) or exercise (value 1).

| id | master_id | user_id | entry_date | note_type | body | uid |
|----|-----------|---------|------------|-----------|------|-----|
| Filter | Filter | Filter | Filter | Filter | Filter | Filter |
| 3 | 10893913 | 1 | 2015-01-16 | 0 | Eat less, live more | 23802195930941 |
| 13 | 10893915 | 1 | 2015-01-16 | 1 | Work hard | 23802195931069 |
| 16 | 10924345 | 2 | 2015-01-23 | 0 | Is this entry visible in the database? | 94179693604669 |
| 17 | 10924348 | 2 | 2015-01-23 | 1 | This is exercise note | 93629946179389 |

**Figure 4. Diary notes of users**

The other eight tables containing artefacts of forensic interest store detailed information about users' fitness and food habits.

## *Findings: RunKeeper*

RunKeeper (FitnessKeeper 2015), considered the best running app by LifeHacker (Klosowski 2012), logs walks, bike rides, hikes and puts all the statistics (pace, distance, and time) at  users' fingertips. It organises the data into charts so that users can track their progress. The app also keeps track of the locations along running routes by letting users share their photos with friends.

### Databases

Forensic analysis indicated that the files and databases generated by RunKeeper (Version 5.4) are stored in the internal device memory which is normally inaccessible by users. For example, the databases of RunKeeper are stored in the `/data/data/com.fitnesskeeper.runkeeper.pro` directory. The names, locations and contents of the artefacts generated by the app are listed in Table 4.

|   | Content | Directory | Files |
|---|---|---|---|
| 1 | Main Database | `/data/data/ fitnesskeeper.runkeeper.pro /databases` | RunKeeper.sqlite (SQLite3, 31 tables) |
| 2 | Images posted by users | `/data/data/ fitnesskeeper.runkeeper.pro /cache/Picasso-cache` | Images are stored as .jpg/.png files |

**Table 4. RunKeeper Artefacts**

The `RunKeeper.sqlite` database is the main database of RunKeeper with 31 tables. Table 5 lists the seven tables containing artefacts of forensic interest.

|   | Table name | Content |
|---|---|---|
| 1 | `deleted_trips` | Trips deleted by user |
| 2 | `feed` | Activities posted by user |
| 3 | `friends` | List of user's friends |
| 4 | `status_updates` | Images uploaded during trips by user |
| 5 | `trip_settings` | User settings for each trip |
| 6 | `points` | Places visited during  all the trips |
| 7 | `trips` | Information about each trip |

**Table 5. Tables in the `RunKeeper.sqlite` database of RunKeeper**

### Trips

The analysis of the trips completed by users can provide investigators important clues about the specific locations travelled by the users. Three tables, `trip_settings`, `trips` and `points`, in the `RunKeeper.sqlite` database of RunKeeper can be analysed to obtain a complete map of user trips.

The `trip_settings` table stores user information about each trip (Table 6). The `setting_key` attributes that are of forensic interest include `name` (in plaintext) and `birthday` (in Unix millisecond epoch time). The app can be used by multiple users on a single device. The `trip_settings` table stores the trip information even after a user has logged out of the device.

| Field name | Meaning |
|---|---|
| `id` | A sequential value or serial number |
| `trip_id` | Unique identifier for each trip |
| `setting_key` | List of attributes |
| `setting_value` | Value for different attributes |

**Table 6. Structure of `trip_settings` table in `RunKeeper.sqlite` database**

Within the `trips` table (Figure 5) , the `_id` field identifies a trip by a numerical value starting from 1. When a user logs out of the device and another user signs in, trips made by the previous user are wiped from the database. For the currently logged in user, the value of the `_id` field starts from the value where the previous user left (e.g. if previous user had five trips, then the new user will have 6 as the value of the first trip in the `_id` field). The `start_date` field specifies the time and date in Unix millisecond epoch time. The `elapsed_time` field records the trip duration in seconds. The `distance` field records the distance of trips in metres. The `calories` field indicates the calorie burnt during the trip. The `notes` field stores user written notes during the trip.

| _id | ext_trip_id | start_date | name | distance | elapsed_time |
|---|---|---|---|---|---|
| Filter | Filter | Filter | Filter | Filter | Filter |
| 1 | 502472481 | 1422297991793.0 | | 19.8604922024573 | 113.0 |
| 2 | 502473487 | 1422298294182.0 | | 3000.0 | 30.0 |
| 3 | 502480002 | 1422299047685.0 | | 19.0393058849443 | 621.0 |
| 4 | 502480917 | 1422299849693.0 | | 0.0 | 5.0 |
| 6 | 0 | 1422362022221.0 | | 207.516353980068 | 219.0 |

**Figure 5. A snapshot of `trips` table in `RunKeeper.sqlite` database**

The `points` table helps an investigator to locate the map coordinates of a user's route. The structure of the table is described in Table 7.

| Field name | Meaning |
|---|---|
| `_id` | A sequential value or serial number |
| `trip_id` | Identifying different trips with different values |
| `latitude` | List of attributes |
| `longitude` | Value for different attributes |
| `altitude` | Sea level altitude of the location |
| `time_interval_at_point` | Time taken in seconds to reach the location |
| `distance_from_last_point` | Distance in metres from last location |
| `distance_at_point` | Distance in metres from the beginning location |

**Table 7. Structure of the `points` table**

For any trip, its value in the `trip_id` field of the `points` table is the same as the `_id` value in the `trips` table. During a trip, a user may travel to various locations. Different locations travelled during a single trip share the same `trip_id` as the `_id` field in the `trips` table. The `latitude`, `longitude` and `altitude` fields represent the geographical position of the user during the trip. The `time_interval_at_point` field represents time elapsed in seconds from the beginning of the trip to the end of the trip (i.e. arriving at the destination). The `distance_from_last_point` and

`distance_at_point` fields represent the distance in metres travelled by the user from the last point and the starting point to the final destination of the trip respectively (Figure 6).

Collectively, the artefacts recovered from the three tables, `trip_settings`, `trips` and `points`, facilitate the reconstruction of locations visited by users, and the time and duration of each trip.

| _id | ext_point_id | trip_id | has_been_sent | latitude | longitude | altitude | time_at_point | time_interval_at_point | speed_from_last_point | distance_from_last_point | point_type | accuracy | distance_at_point |
|---|---|---|---|---|---|---|---|---|---|---|---|---|---|
| Filter | Filter | Filter | Filter | Filter | Filter | Filter | Filter | Filter | Filter | Filter | Filter | Filter | Filter |
| 5 | 1 | 0 | | -34.8067581 | 138.61073062 | 11.0 | 0 | 0.0 | 0.0 | 0.0 | 0 | 34.0 | 0.0 |
| 6 | 1 | 0 | | -34.80676169 | 138.61074841 | 11.0 | 0 | 3.0 | 0.55752168124017... | 1.67256504372052 | 4 | 15.0 | 1.67256504372052 |
| 7 | 1 | 0 | | -34.80681559 | 138.61082237 | 11.0 | 0 | 41.0 | 0.22021288270530... | 9.02872819091741 | 4 | 4.0 | 10.70129323463... |
| 8 | 1 | 0 | | -34.80684196 | 138.61091741 | 11.0 | 0 | 88.0 | 0.10408180645249... | 9.15919896781941 | 5 | 4.0 | 19.86049220245... |
| 13 | 3 | 0 | | -34.80677696 | 138.6107484 | 11.0 | 0 | 0.0 | 0.0 | 0.0 | 0 | 18.0 | 0.0 |
| 14 | 3 | 0 | | -34.80677456 | 138.61074602 | 11.0 | 0 | 2.0 | 0.16219700611194... | 0.32439401222389... | 4 | 10.0 | 0.32439401222... |
| 15 | 3 | 0 | | -34.80681168 | 138.61084043 | 11.0 | 0 | 162.0 | 0.05899340893580... | 9.55693224760027 | 4 | 11.0 | 9.88132625982416 |
| 16 | 3 | 0 | | -34.80683205 | 138.61093762 | 11.0 | 0 | 167.0 | 0.05483830134802... | 9.15797962512019 | 5 | 11.0 | 19.03930598849444 |
| 60 | 6 | 0 | | -34.81003651 | 138.61946183 | 17.0 | 0 | 0.0 | 0.0 | 0.0 | 0 | 25.0 | 0.0 |
| 61 | 6 | 0 | | -34.81007723 | 138.61959405 | 17.0 | 0 | 44.0 | 0.29475996766732... | 12.9694385773623 | 4 | 15.0 | 12.9694385773623 |
| 62 | 6 | 0 | | -34.81006891 | 138.61948439 | 17.0 | 0 | 55.0 | 0.16778851086655... | 9.22726809766029 | 4 | 5.0 | 22.19670667502... |

**Figure 6. Artefacts found in `points` table**

### Other tables

The remaining tables containing artefacts of forensic interest are `deleted_trips`, `feeds`, `friends` and `status_updates` tables.

The `deleted_trips` table stores information of trips deleted by users. The `_id` field of the `deleted_trips` table contains the sequence number of the trip. The location entries from the `points` table are removed for a deleted trip. An investigator can only obtain the time of deletion for the trips without additional information about the locations travelled during the trips.

The `feeds` table contains the news feeds posted by users, and the corresponding timestamp in Unix millisecond epoch time, as well as user ID, username, and JavaScript Object Notation (JSON) data containing the duration, notes, distance covered during the trip.

The `friends` table stores the names and email IDs of users' friends in plaintext. The table also stores the time when each friend was active. The `status_updates` table stores the timestamp, latitude and longitude along with the trip IDs when an image was uploaded by a user.

| _id | name | fbuid | email | status | currentMonthActivityCount | lastActive |
|---|---|---|---|---|---|---|
| Filter | Filter | Filter | Filter | Filter | Filter | Filter |
| 39876746 | A▓▓▓▓▓ar | 0 | al▓▓▓▓▓ar@g▓▓l.com | 1 | 2 | 1422304983101 |

**Figure 7. Contents of `friends` table**

### *Findings: Period Calendar*

Period Calendar is currently the highest rated (4.7) menstrual cycle calculator in Android (Abishkking 2015). The app is rated among the top five apps in mHealth category in over 49 countries with over 30 million Android users. The app tracks periods, moods and symptoms.

**Databases**

Our forensic analysis indicated that the files and databases generated by Period Calendar (Version 1.51) are stored in the internal device memory which is normally inaccessible by users. For example, the databases of Period Calendar are stored in the `/data/data/com.popularapp.periodcalendar/databases` directory. The names, locations and contents of the artefacts generated by the app are listed in Table 8.

|   | Content | Directory | Files |
|---|---------|-----------|-------|
| 1 | Main Database | `/data/data/ fitnesskeeper.runkeeper.pro /databases` | PC.db (SQLite3, 5 tables) |
| 2 | Pill Database | `/data/data/ fitnesskeeper.runkeeper.pro /databases` | PC_PILL.db (SQLite3, 4 tables) |
| 3 | Backup Databases | `/data/data/ fitnesskeeper.runkeeper.pro /app_Backup_db` | PC.db (SQLite3, 5 tables)<br><br>PC_PILL.db (SQLite3, 4 tables) |

**Table 8.Period Calendar Artefacts**

The `PC.db` database is the main database of Period Calendar with five tables, and Table 9 lists the three tables that contain artefacts of forensic interest.

|   | Table name | Content |
|---|-----------|---------|
| 1 | `user` | List of the users with passwords |
| 2 | `note` | Diary notes inserted by users |
| 3 | `period` | Period start time and length of users |

**Table 9. Tables in the `PC.db` database of Period Calendar app**

The `PC_PILL.db` database contains four tables, and only two tables were determined to contain artefacts of forensic interest – see Table 10.

|   | Table name | Content |
|---|-----------|---------|
| 1 | `pill` | Pills used by users including date and time |
| 2 | `pill_record` | Details about the pills |

**Table 10. Tables in the `PC_PILL.db` database of Period Calendar app**

**User Table**

The `user` table in the `pc.db` database stores records of users. The structure of the `user` table is given in Table 11.

| Field name | Meaning |
|---|---|
| `_id` | A sequential value or serial number |
| `uid` | Numerical values Identifying different users |
| `username` | List of user attributes |
| `password` | Plaintext passwords of different users |
| `email` | Email address of the users |
| `question` | Secret question to retrieve password |
| `answer` | Answer of the secret question in plaintext |
| `setting` | User height, weight, period lengths, temperature, pill lists |
| `temp1` | Unused fields |
| `temp2` | |
| `temp3` | |

**Table 11. Structure of `user` table of Period Calendar**

The app can be used by multiple users on a single device, and each user has a unique identifier (see `uid` field in Figure 8). Usernames are stored in the `user_name` field in plaintext. Similarly, user's passwords and email IDs are stored as plaintext in the `password` and `email` fields respectively. The `question` and `answer` fields store users' secret questions and corresponding answers in plaintexts, which are used to retrieve forgotten user passwords. The `setting` field stores different user attributes such as user height, weight, temperature, period lengths, and pill lists.

**Figure 8. Plaintext passwords, secret questions and answers in `user` table**

**Other tables**

The `note` table stores remaining information about the users including notes, pills taken, temperature, weight, symptoms and moods – see Table 12.

| Field name | Meaning |
|---|---|
| `_id` | A sequential value or serial number |
| `uid` | Numeric values identifying different users |
| `date` | Timestamp in Unix millisecond epoch |
| `note` | User notes |
| `pill` | Names of pills taken by a user |
| `temperature` | Body temperature of a user in degree Celsius |
| `weight` | Weight of a user in lb |
| `symptoms` | Different symptoms and moods of users |
| `moods` | |

**Table 12. Structure of `note` table of Period Calendar app**

The `period` table stores the period start time and length for different users in Unix millisecond epoch. Different users have different `uid` values (Figure 9). The `pill` table in the `PC_PILL.db` database stores the name of the pills taken by the users along with the corresponding information(i.e. when a pill was first taken and the time of the next medication).

| _id | menses_start | menses_length | period_length | pregnancy | uid |
|---|---|---|---|---|---|
| Filter | Filter | Filter | Filter | Filter | Filter |
| 4 | 1421587800000 | 3 | 19 | 0 | 0 |
| 6 | 1419255000000 | -3 | 27 | 0 | 0 |
| 7 | 1423229400000 | -3 | 27 | 0 | 0 |
| 8 | 1422279000000 | -4 | 26 | 0 | 1 |
| 9 | 1422711000000 | 6 | 30 | 0 | 2 |

**Figure 9. Snapshot of the `period` table**

# Proposed Forensic Taxonomy for mHealth Apps

The increasing number of mHealth apps makes it very difficult and challenging to construct a forensic taxonomy for all existing mHealth apps. Therefore, the taxonomy is constructed based on our study of the 40 apps studied in the preceding section.

## *mHealth App Categories*

Ozdalga et al. (2012) proposed four categories for mHealth apps:

*Patient care and monitoring*: This category includes the mHealth apps which use mobile devices and global positioning system (GPS) to remotely monitor patients. For example, a sensor fitted in the shoes of a patient who has suffered from recent strokes can be used to communicate with the mobile apps to track the patient' activities (Edgar et al. 2010).

*Health apps for the layperson*: Weight loss (e.g. MyFitnessPal) and fitness (e.g. RunKeeper) apps are among the most popular apps in this category.

*Communication, education, and research*: These apps allow users to receive information in a timely manner and be more efficient during critical situations (e.g. natural disasters and medical emergencies). For example, a number of hospitals (e.g. Doylestown Hospital and George Washington University Hospital) have developed several projects to give physicians secure access to medical databases using smartphones (Hamou et al. 2010).

*Physician or student reference apps*: These apps are designed to facilitate better decision making process and to reduce medical errors. For example, Medscape, a widely used app by  physicians, provides a large, easy to use index of drug reference tools, as well as medical news updates.

### Forensic Artefact Categories

From the artefacts determined from our forensic analysis of the 40 most popular Android mHealth apps, we categorised the artefacts into seven groups:

*Databases*: Most of the Android apps generate their own databases in the internal device memory, where the latter is normally inaccessible by users. These databases generally provide references that can be used to locate useful user information.

*User credentials*: Apps may require users to login using their user credentials (e.g. username and password, PIN, and authentication tokens) in order to use the apps. Therefore, user credentials should be an artefact that forensic investigators seek to locate during the app forensic process (e.g. determine whether the credentials are stored in and can be recovered from the app's databases).

*User personal details*: User personal details include name, gender, date of birth, email address, height, weight and other personal data would be helpful for forensic investigators to positively identify the app or device users.

*User activities*: The mHealth apps require  users to enter their day-to-day food habit, health conditions, activity or exercise details, diagnosis details, medication details and symptom details, etc.

*User location*: Fitness apps allow users to keep track of their exercise, running, jogging, cycling and other activities. These apps generally store the geographical coordinates of the user location during these activities which can provide useful evidence to the investigators.

*Activity timestamps*: Another important artefact is the timestamp of the user activity. For example, linking activity timestamps with corresponding user locations (e.g. geographical coordinates) and other relevant information (e.g. CCTV feeds) would provide useful information in an investigation.

*Images*: This artefact includes profile images, and images taken and posted from a location.

### The Forensic Taxonomy Model

Our proposed mHealth forensic taxonomy has two dimensions. The mHealth app categories are represented in one dimension, and the forensic artefacts are represented in the other dimension (Figure 10). We summarised the findings of the 40 case study apps using the two-dimensional taxonomy in Appendix A.

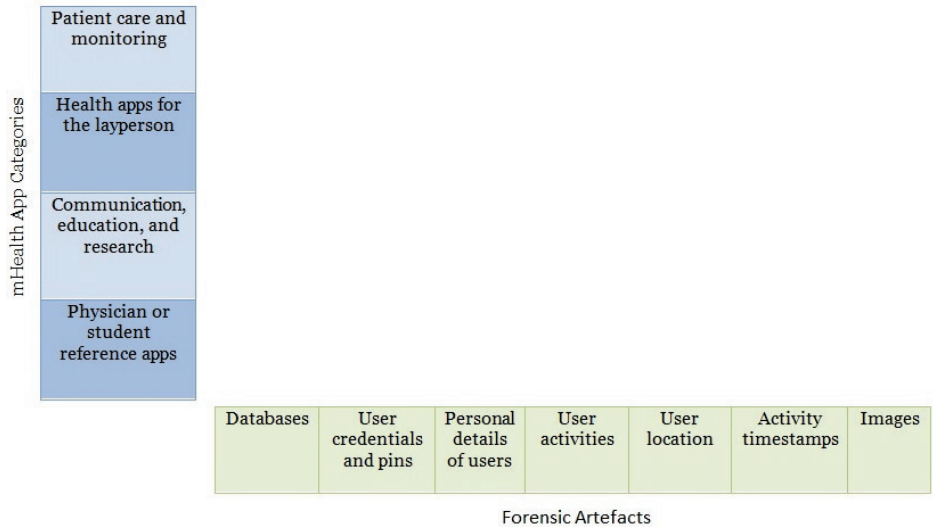

**Figure 10. A two-dimensional mHealth forensic taxonomy model**

## Concluding remarks

In this paper, we analysed 40 popular mHealth apps downloaded from Google Play store. The findings of our analysis are summarised in Table 13.

We found that the majority of the Health apps for the layperson stored personal user details, user activities and activity timestamps in the databases. The patient care and monitoring apps and the communication, education, and research apps tend to store information relating to user activities, activity timestamps and user images in their databases.

Findings from this research will be of importance to the forensic community, as well as in criminal investigations and civil litigation matters involving the 40 apps examined in this paper. The user activities, their location information and timestamps can facilitate the reconstruction of a user's whereabouts.

Findings are accurate at the time of this research, but new releases of mHealth apps may change the way data are stored on the devices, as well as the type of data that can be forensically recovered from the devices. Therefore, future work would include examining other and new releases of mHealth apps, and potentially include additional artefact categories to the proposed taxonomy.

| App ID | App Name | Version | App Category | | | | Artefact Category | | | | | | |
|---|---|---|---|---|---|---|---|---|---|---|---|---|---|
| | | | Patient care and monitoring | Health apps for the layperson | Communication, education, and research | Physician or student reference | Databases | User credentials and pins | Personal details of users | User activities | User location | Activity timestamps | Images |
| App1 | MyFitnessPal | 3.6.1 | P | F | P | N | F | P | F | F | N | F | F |
| App2 | RunKeeper - GPS | 5.4 | N | F | N | N | F | N | N | F | F | F | N |
| App3 | Period Calendar | 1.51 | P | F | N | N | F | F | F | F | N | P | N |
| App4 | WebMD | 3.5 | N | F | F | P | P | N | N | P | N | N | N |
| App5 | Blood Pressure (BP) Watch | 3.0.11 | P | F | N | N | F | N | P | F | N | F | N |
| App6 | Calorie Counter by FatSecret | - | P | F | P | N | F | N | N | F | N | P | N |
| App7 | Google Fit | 1.51.07 | N | F | N | N | F | N | N | F | N | F | N |
| App8 | MyNetDiary Calorie Counter PRO | 2.2.0 | P | F | P | N | N | N | N | N | N | N | F |
| App9 | Drugs.com Medication Guide | 1.23 | N | F | F | P | F | N | F | N | N | P | N |
| App10 | My Diet Diary Calorie Counter | 1.9.11 | P | F | P | N | F | N | P | F | N | F | N |
| App11 | Calories! Basic – cal counter | 1.1.7 | P | F | P | N | F | N | N | F | N | F | N |
| App12 | Period Tracker | 2.0.6.4 | P | F | N | N | F | N | N | F | N | N | N |
| App13 | Calorie Counter | 4.2.5 | P | F | P | N | F | N | F | F | N | F | N |
| App14 | My Pregnancy Today | 1.14.0 | P | F | N | N | N | P | N | N | N | N | F |
| App15 | Water Your Body | 3.062 | N | F | N | N | F | N | N | N | N | N | N |
| App16 | Instant Heart Rate | - | P | F | N | N | F | N | N | N | N | N | N |
| App17 | Calm – Meditate, Sleep, Relax | 1.9.4 | N | P | N | N | F | N | F | N | N | F | N |
| App18 | Runtastic Pedometer | 1.5.1 | N | F | N | N | F | N | N | N | N | N | N |
| App19 | Smiling Mind | 2.0.3 | N | F | N | N | F | N | F | N | N | N | N |
| App20 | Pedometer | 5.10 | N | F | N | N | F | N | N | N | N | N | N |
| App21 | Quit Now: My QuitBuddy | 2.1 | P | F | N | N | N | N | N | N | N | N | F |

| | App Name | Version | App Category | | | | Artefact Category | | | | | | |
|---|---|---|---|---|---|---|---|---|---|---|---|---|---|
| | | | Patient care and monitoring | Health apps for the layperson | Communication, education, and research | Physician or student reference | Databases | User credentials and pins | Personal details of users | User activities | User location | Activity timestamps | Images |
| App22 | Mindbody Connect | 2.8.3 | N | P | N | N | F | N | N | P | P | N | N |
| App23 | My Baby Today | - | N | F | N | N | F | N | N | N | N | P | N |
| App24 | Lifesum-Calorie Counter | - | P | F | P | N | F | N | P | F | N | F | F |
| App25 | Quit Smoking – QuitNow! | - | N | F | N | N | N | N | N | N | N | N | N |
| App26 | Strava Running and Cycling GPS | 4.3.1 | N | F | N | N | F | N | F | F | F | F | N |
| App27 | Lorna Jane | 1.2 | N | F | N | N | F | N | N | P | F | F | F |
| App28 | Walk with Map My Walk | 3.5.1 | N | F | N | N | F | N | F | F | F | F | P |
| App29 | FitNotes – gym Workout Log | 1.12.0 | P | F | N | N | F | N | N | F | N | P | N |
| App30 | Nike+ Running | 1.5.2 | N | F | N | N | F | N | F | F | N | F | F |
| App31 | 30 day Ab Challenge | 2.0 | N | P | N | N | N | N | N | N | N | N | N |
| App32 | Genesis YNB | 1.0.2 | N | F | N | N | F | N | F | N | P | N | F |
| App33 | BMI Calculator | - | N | P | N | N | N | N | N | N | N | N | N |
| App34 | Endomondo Running Cycling Walking | 10.6.3 | N | F | N | N | F | N | N | F | F | F | F |
| App35 | Fitness Buddy: 300+ Exercises | 3.10 | N | F | N | N | F | N | N | F | N | N | F |
| App36 | My Tracks | 2.0.9 | N | F | N | N | F | N | N | F | F | F | F |
| App37 | Under Armour Record | 2.1.1 | N | F | N | N | F | N | F | F | N | F | F |
| App38 | Noom Walk Pedometer: Fitness | 1.1.0 | N | F | N | N | F | N | N | F | F | F | F |
| App39 | Bleep Fitness Test | 1.8 | N | P | N | N | F | N | F | F | N | P | N |
| App40 | BodySpace-Social Fitness | 1.3.9 | N | F | N | N | F | N | F | F | N | F | F |

**Table 13. Summary of findings**

Notes: "F" - detailed information was recovered; "P" - only partial information was recovered (e.g. artefacts from some apps provided partial timestamp such as only the date of the activity rather than the time in hours, minutes and seconds);and "N" - unsupported category.

# APPENDIX A

## *Forensic Taxonomy of the mHealth Apps*

| App Category/ Artefact Category | | Databases | | User credentials | | Personal details of users | | User activities | | User location | | Activity timestamps | | Images | |
|---|---|---|---|---|---|---|---|---|---|---|---|---|---|---|---|
| | | Full | Partial | Full | Partial | Full | Partial | Full | Partial | Full | Partial | Full | Partial | Full | Partial |
| Patient care and monitoring | Full | | | | | | | | | | | | | | |
| | Partial | App1 App3 App5 App6 App10 App11 App12 App13 App16 App24 | | App3 | App1 App14 | App1 App3 App13 | App5 App10 App24 | App1 App3 App5 App6 App10 App12 App13 App24 | App11 | | | App1 App5 App10 App11 App13 App24 | App3 App6 App12 | App1 App8 App14 App21 App24 | |
| Health apps for the layperson | Full | App1 App2 App3 App5 App6 App9 App10 App11 App12 App13 App15 App16 App18 App19 App23 App24 App26 App27 App28 App29 App30 App34 App35 App36 App40 | App4 App7 App20 | App3 | App1 App14 | App1 App3 App9 App13 App19 App23 App26 App28 App30 App32 App37 App40 | App5 App10 App24 | App1 App2 App3 App5 App6 App10 App12 App13 App15 App18 App20 App24 App26 App28 App29 App30 App34 App35 App36 App37 App40 | App4 App7 App11 App27 | App2 App26 App27 App28 App34 App36 | App32 | App1 App2 App5 App7 App10 App11 App13 App18 App19 App20 App24 App26 App27 App28 App30 App35 App35 App37 App37 | App3 App6 App9 App12 App23 App29 App40 | App1 App8 App14 App21 App24 App27 App30 App34 App37 App40 | |

| App Category/ Artefact Category | | Databases | | User credentials | | Personal details of users | | User activities | | User location | | Activity timestamps | | Images | |
|---|---|---|---|---|---|---|---|---|---|---|---|---|---|---|---|
| | | Full | Partial | Full | Partial | Full | Partial | Full | Partial | Full | Partial | Full | Partial | Full | Partial |
| Health apps for the layperson | Partial | App17 App22 App38 App39 | | | | App17 App39 | | App38 App39 | App22 | | App22 | App17 App38 | App39 | App38 | |
| Communication, education, and research | Full | App9 | App4 | | | App9 | | | App4 | | | | App9 | | |
| Communication, education, and research | Partial | App1 App6 App10 App11 App13 App24 | | | App1 | App1 App13 | App10 App24 | App1 App6 App10 App13 App24 | App11 | | | App1, App10 App11 App13 App24 | App6 | App1 App8 App24 | |
| Physician or student reference | Full | | | | | | | | | | | | | | |
| Physician or student reference | Partial | App9 | App4 | | | App9 | | | App4 | | App9 | | | | |